\documentclass[10pt,twocolumn,twoside,submit]{JCNtran}

\usepackage{amsmath}
\usepackage{epsfig}
\usepackage[latin1]{inputenc}
\usepackage[T1]{fontenc}
\usepackage{balance}

\def\BibTeX{{\rm B\kern-.05em{\sc i\kern-.025em b}\kern-.08em
    T\kern-.1667em\lower.7ex\hbox{E}\kern-.125emX}}

\begin{document}
\bibliographystyle{jcn}

\title{Computationally Efficient Implementation of a Hamming Code Decoder using Graphics Processing Unit}

\author{Md Shohidul Islam, Cheol-Hong Kim,
and Jong-Myon Kim\textsuperscript{*}
\thanks{Manuscript received Feb. 5, 2014: approved for publication by Prof. Ali Abedi, Dec. 1, 2014.
This work was supported by a grant from the National Research Foundation of Korea (NRF) funded by the Korean government (MEST) (No. NRF-2013R1A2A2A05004566).}
\thanks{Md Shohidul Islam and Jong-Myon Kim are with the School of Electrical Engineering, University of Ulsan, Korea, email: shohid@mail.ulsan.ac.kr,jmkim07@ulsan.ac.kr. *Jong-Myon Kim is a corresponding author.}
\thanks{Cheol-Hong Kim is with the School of Electronics and Computer Engineering, Chonnam National University, Gwangju, Korea, email: chkim22@chonnam.ac.kr. }
} 
\markboth{JOURNAL OF
COMMUNICATIONS AND NETWORKS}{Shohidul \lowercase{\textit{et al}}.: Computationally Efficient Implementation of a Hamming Code Decoder...} \maketitle

\begin{abstract}
This paper presents a computationally efficient implementation of a Hamming code decoder on a graphics processing unit (GPU) to support real-time software-defined radio (SDR), which is a software alternative for realizing wireless communication. The Hamming code algorithm is challenging to parallelize effectively on a GPU because it works on sparsely located data items with several conditional statements, leading to non-coalesced, long latency, global memory access, and huge thread divergence. To address these issues, we propose an optimized implementation of the Hamming code on the GPU to exploit the higher parallelism inherent in the algorithm. Experimental results using a compute unified device architecture (CUDA)-enabled NVIDIA GeForce GTX 560, including 335 cores, revealed that the proposed approach achieved a 99x speedup versus the equivalent CPU-based implementation. 
\end{abstract}


\section{\uppercase{Introduction}}
\label{sec:introd}

Many existing wireless communication systems employed application specific integrated circuits (ASICs) based dedicated devices for particular communication protocol standards, including worldwide interoperability for microwave access (WiMAX, IEEE 802.16), Wi-Fi (IEEE802.11), digital high definition TV, wideband code division multiple access (W-CDMA), and global system for mobile communication (GSM)~[1]-[7].
However, the fixed functionality of such ASIC devices limits their application to emerging communication standards because they were fixed for specific coding schemes, data rates, frequency ranges, and types of modulation~[8]. 
In addition, manufacturing costs are high and time-to-market of hardware devices is long~[9]. 

Software-defined radio (SDR) is an emerging technology that offers software alternatives to existing hardware solutions for wireless communication~[10],[11], 
SDR technology has recently attracted the interest of the communication research community ~[12],[13]. 
SDR comprises software implementation of multi-standard and multi-protocol communication systems using one hardware platform~[14].
It allows system reconfiguration by using software commands, because users are required to switch from one standard to another standard very frequently~[15]. 
In addition, it enables the radio device to change transmitting and receiving characteristics by means of the software without altering the hardware platform~[15]. 
In SDR, some or all of the physical layer functions are coded in the software, which runs on general-purpose programmable processors (GPPs) and digital signal processors (DSPs) ~[16],[17]. 
GPPs and DSPs offer the necessary programmability and flexibility for various SDR applications. However, neither GPPs nor DSPs can meet the much higher levels of performance required by high computational workloads in SDR~[18].

Among many available computational models, graphics processing units (GPUs) perform well when performing latency-tolerant, highly parallel, and independent tasks. Attracted by the features of modern GPUs, many researchers have developed GPU-based SDR systems including turbo decoders, LDPC decoders, Viterbi decoders, and MIMO detectors to meet the high throughput required by the SDR algorithm~[19]-[25].
In this paper, we present an optimized implementation of a Hamming decoder on a GPU; the Hamming decoder is widely used as a forward error correction (FEC) mechanism in wireless communication. Practical applications of the Hamming decoder include Ethernet (IEEE 802.3), WiMAX (IEEE 802.16e), Wi-Fi (IEEE 802.11n), telecommunication, digital video broadcasting-satellite second generation (DVB-S2), wireless sensor networks (WSNs), underwater wireless sensor networks (UWSNs), and space communication~[26]-[31].

Contributions of this study are as follows:
\begin{itemize}
\item[\textbullet] This paper presents a massively parallel and optimized implementation of a Hamming decoder on a GPU by exploring memory transfer, memory transaction, and kernel computation.
\item[\textbullet ]The performance of the Hamming decoder on the GPU is thoroughly evaluated for various packet sizes, code lengths, and error tolerance.
\item[\textbullet] The performance of the proposed GPU approach is compared with the equivalent sequential approach run on a conventional CPU.
\end{itemize}

The remainder of this paper is organized as follows. A review of the Hamming decoder is provided in Section~II
, optimization and GPU implementation of the Hamming decoder are presented in Section~III
, and experimental results and analysis are discussed in Section~IV.

\vspace{10pt}
\section{\uppercase{Review of the Hamming Decoder}}
\label{sec:implem}

Hamming decoding is performed at the destination end of the packet, and involves the exact reverse process of encoding performed at the transmitter end. Figure 1 shows a Hamming code decoder that consists of three components: splitter, decoder, and merger. The splitter receives the Hamming encoded packet,  \textbf{H}=$\{$\textbf{b}; \textbf{b}=0 | \textbf{b}=1$\}$, and splits the message into t segments, $\textbf{H}_{1}$, $\textbf{H}_{2}$,$...$, $\textbf{H}_{t}$, where \textit{t} is the error tolerance. The main decoder consists of three fundamental units: error detection (ED), error correction (EC), and redundancy remover (RR). We use the terms $\textit{packet}$ and $\textit{message}$ interchangeably throughout this paper.

\begin{figure}[h]
\begin{center}
\epsfxsize=8.2cm \leavevmode\epsfbox{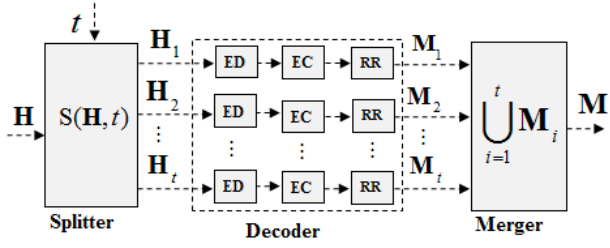}
\caption{Components in a Hamming code decoder.} \label{fig:numbytes1}
\end{center}
\end{figure}

In the encoding process, some redundancy or checksum bits are incorporated along with the original message for the purpose of error detection, and these should be removed once they have served their purpose. Subsequently, the decoder retrieves message segments $\textbf{M}_{1}$, $\textbf{M}_{2}$,..., $\textbf{M}_{t}$, and the merger unifies them to produce the decoded packet, \textbf{M}, which is similar to the original packet sent by the sender or transmitter. 

\vspace{10pt}
\section{\uppercase{GPU-based implementation of the Hamming decoder}}
\label{sec:gpubased}

This section presents a computationally efficient implementation of the Hamming code algorithm on a GPU. Encoded packets, namely $\textit{P}_{1}$, $\textit{P}_{2}$,$...$,$\textit{P}_{n}$, are primarily received in the receiver buffer and the entire task can be divided into three steps, as shown in Figure 2: (i) pre-processing in the CPU, (ii) packet transfer between the CPU and GPU, and (iii) device kernel execution (DKE). All of these steps are performed in an optimized manner from a GPU computing viewpoint.

\begin{figure}[h]
\begin{center}
\epsfxsize=7.75cm \leavevmode\epsfbox{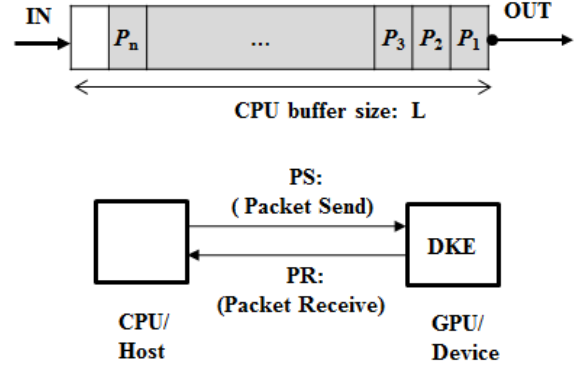}
\caption{Task partitioning for the proposed Hamming decoder.} \label{fig:numbytes1}
\end{center}
\end{figure}

Figure 3 depicts an execution flow of the entire decoding process in the destination end of a network data packet, where regular blocks represent the steps executed on the CPU and the dotted blocks represent the tasks in the GPU. The CPU and GPU are also called the Host and Device, respectively. At the outset, the encoded packet, \textbf{H}, undergoes pre-processing in the CPU, which is explained in Section 3.1, before being transferred to the GPU. 
A parallel algorithm executed on the GPU is called a kernel, and the proposed approach configures two kernels, namely \textit{checksum} and \textit{error}, as indicated in Figure 3. The \textit{checksum} kernel computes the redundancy information, and the \textit{error} kernel performs error detection, correction, redundancy removal, and finally retrieves the original packet. This packet, now referred to as the decoded packet, is transferred from the GPU back to the CPU.

\begin{figure}[h]
\begin{center}
\epsfxsize=5cm \leavevmode\epsfbox{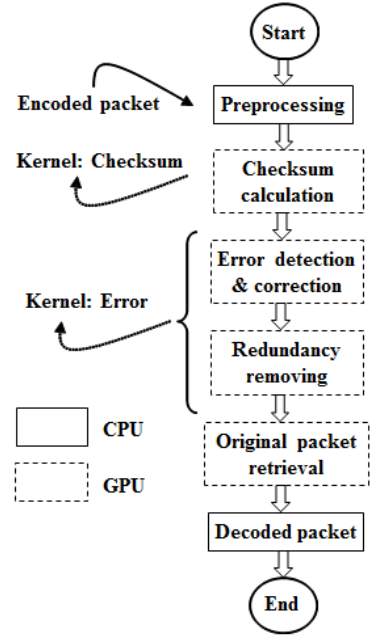}
\caption{Execution flow of the Hamming decoding procedure on a CPU and GPU.} \label{fig:numbytes1}
\end{center}
\end{figure}

\vspace{10pt}
\subsection{Packet pre-processing}
\label{sec:packetpre}

Instead of transferring the encoded packet, \textbf{H}= $\textbf{H}_{1}$+$\textbf{H}_{2}$+$...$+$\textbf{H}_{t}$, to the GPU immediately, packet pre-processing is first performed, because the first step in a GPU is to calculate checksums on each of $\textbf{H}_{1}$,$\textbf{H}_{2}$,$...$,$\textbf{H}_{t}$ ; this process accesses sparsely located elements in global memory. Considering any $\textbf{H}_{i}$, the index sets $\textbf{I}_{0}$, $\textbf{I}_{1}$,$...$,$\textbf{I}_{\textbf{|}\textbf{R}\textbf{|}-1}$, shown in Figure 4, access indices of $\textbf{H}_{i}$ that are not completely adjacent. For instance, (7, 4) Hamming code has |$\textbf{H}_{i}$|=7+4=11 bits and |\textbf{R}|=4. Thus, $\textbf{I}_{0}$=$\{$1,3,5,7,9,11$\}$, $\textbf{I}_{1}$=$\{$2,3,6,7,10,11$\}$, $\textbf{I}_{2}$=$\{$4,5,6,7$\}$, and $\textbf{I}_{3}$=$\{$8,9,10,11$\}$. Consequently, data items accessed by $\textbf{I}_{0}$ and $\textbf{I}_{1}$are clearly non-adjacent. Even though elements of $\textbf{I}_{2}$ and $\textbf{I}_{3}$ apparently seem to be adjacent for a small code length, they include non-adjacent indices for larger values of |$\textbf{H}_{i}$|.

\begin{figure}[h]
\begin{center}
\epsfxsize=8cm \leavevmode\epsfbox{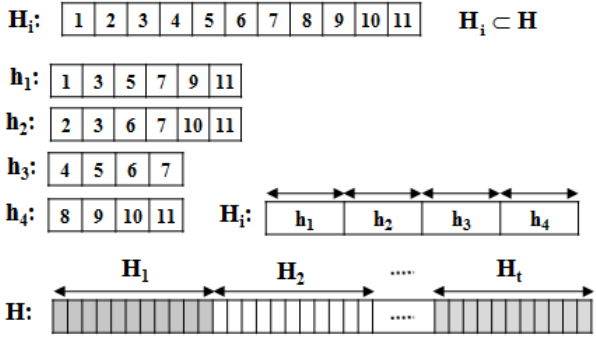}
\caption{Encoded packet re-construction for achieving coalesced global memory access in GPU.} \label{fig:numbytes1}
\end{center}
\end{figure}

If the packet segment is transferred to the GPU without pre-processing, there are two major performance bottlenecks: (a) non-adjacent memory transactions from GPU global memory result in long latency for memory READ, and (b) a number of conditional statements are required to access those locations, leading to thread divergence. These issues are addressed by pre-processing, which achieves coalesced global memory access. To this end, we re-organize the data items of the message segment, $\textbf{H}_{i}$, by arranging those sparse items together and forming clusters such as $\textbf{h}_{1}$, $\textbf{h}_{2}$, $\textbf{h}_{3}$, $\textbf{h}_{4}$ for each group. The clusters are placed side-by-side to shape the new $\textbf{H}_{i}$ as shown in Figure 4 Finally, the reformed encoded packet, \textbf{H}, is created by concatenation of $\textbf{H}_{i}$ $^{\textquoteright}$s such that \textit{i}=1,2,$...$,\textit{t}, and this reformed encoded packet is transferred to the GPU.

\vspace{10pt}
\subsection{Packet transfer between CPU and GPU}
\label{sec:packettra}

Data transfer between host and device is a vital issue in GPU computing. We utilize an optimized data transfer approach to achieve high performance. A GPU facilitates two modes of data transfer: synchronous data transfer (SDT) and asynchronous data transfer (ADT). Referring to Figure 2, three independent tasks, namely encoded packet transfer to the GPU [PS], DKE, and decoded packet receive from the GPU [PR], are accomplished in the GPU. As shown in Figure 5, these tasks are performed concurrently in ADT, where most of the transfer time is hidden by kernel execution. In contrast, SDT takes a long time and completes tasks in a sequential manner. As a result, ADT outperforms SDT by due to its pipelined execution pattern; therefore, we utilize ADT. 

Figure 5 depicts the key differences between SDT and ADT in terms of execution time line. $T_{PS}$, $T_{DKE}$, and $T_{PR}$  indicate the time required for the PS, DKE, and PR, respectively. The SDT based approach takes 3$\times$$T_{PS}$ + 3$\times$$T_{DKE}$+3$\times$$T_{PR}$ time units to process three packets, whereas ADT requires $T_{PS}$ + 3$\times$$T_{DKE}$ + $T_{PR}$. Consequently, ADT saves 2($T_{PS}$ + $T_{PR}$) time units.
In general, the speedup of ADT over SDT for the processing of N packets can be expressed by

\begin{equation}
\begin{aligned}
ADT\,Speedup=\frac{SDT\,Execution\,Time}{ADT\,Execution\,Time}
\end{aligned}
\end{equation}

\begin{equation*}
$$
\[ = \frac{N \times(T_{PS}+T_{DKE}+T_{PR})}{T_{PS}+N \times T_{DKE}+T_{PR}}\]
\[ \approx (N,...,2); \\ when (T_{PS}+T_{PR} \leq T_{DKE},...,T_{PS}+T_{PR} = T_{DKE})   \]
$$
\end{equation*}

Therefore, the minimum expected gain is two times; in our implementation, $T_{PS}$ + $T_{PR}$ $\geq$
 $T_{DKE}$, which accelerates the process toward N.

\begin{figure}[h]
\begin{center}
\epsfxsize=8cm \leavevmode\epsfbox{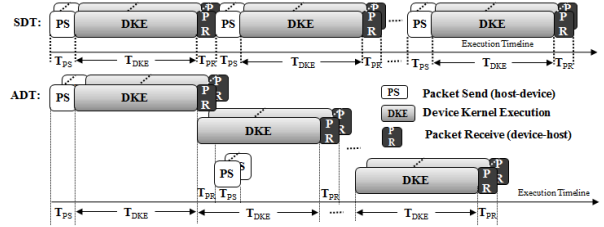}
\caption{Packet transfer using ADT and SDT.} \label{fig:numbytes1}
\end{center}
\end{figure}

\vspace{10pt}
\subsection{Device Kernel Execution (DKE)}
\label{sec:devicekernel}

Algorithms that are executed in parallel on a GPU are called kernels. Algorithm 1 and Figure 6 show a \textit{checksum} kernel that accesses the pre-processed encoded packet in the GPU global memory. The main task of this kernel is to calculate a checksum vector, $\textbf{C}_{i}$, on $\textbf{H}_{i}$, which is the major computation of the Hamming decoder. 

\begin{table}[h]
\label{tab:algorithm1}
\begin{center}
\begin{tabular}{l}
\hline
Algorithm 1: $Kernel-\textit{Checksum} (\textbf{H}_{1}, \textbf{C}_{1})$\\
\hline
\textbf{Input}: Packet segment, $\textbf{H}_{1}$, from the pre-processed\\ encoded packet, \textbf{H}\\
\textbf{Output}: Checksum vector, $\textbf{C}_{1}$\\
\textbf{Step 1}: READ $\textbf{H}_{1}$ from global memory\\
\textbf{Step 2}: For each block of GPU in parallel, do segmentation\\ on $\textbf{H}_{1}$ by
          index set $\textbf{I}_{0}, \textbf{I}_{1},..., \textbf{I}_{\textit{r}(\textbf{H}_{1})}$\\
\textbf{Step 3}: WRITE segments to the shared memory \\ of each block\\
\textbf{Step 4}: Perform module 2 (XOR) operation on the shared \\ memory packet 
          segment\\
\textbf{Step 5}: WRITE the result of Step 4, the checksum vector \\ $\textbf{C}_{1}[1],  
          \textbf{C}_{1}[2],..., \textbf{C}_{1}$[\textit{r}($\textbf{H}_{1}$)], in global memory \\ \hline
\end{tabular}
\end{center}
\end{table}

The notation \textit{r}($\textbf{H}_{i}$) corresponds to the number of redundant or checksum bits required for error detection in the $\textit{i}^{th}$ message segment, $\textbf{H}_{i}$. The kernel declares \textit{r}($\textbf{H}_{i}$) blocks, where each block is responsible for calculating one checksum bit. The checksum calculation is done by a modulo 2 operation on the bit strings stored in each block's shared memory by applying equations, as shown in Figure 6. The proposed approach removes bank conflicts by accessing items from a different shared memory bank. This kernel is accelerated by three optimizations: (i) coalesced global memory access, (ii) bank conflict avoidance by a reduction tree, and (iii) most computations based on shared memory.

\begin{figure*}[t]
\begin{center}
\epsfxsize=12cm \leavevmode\epsfbox{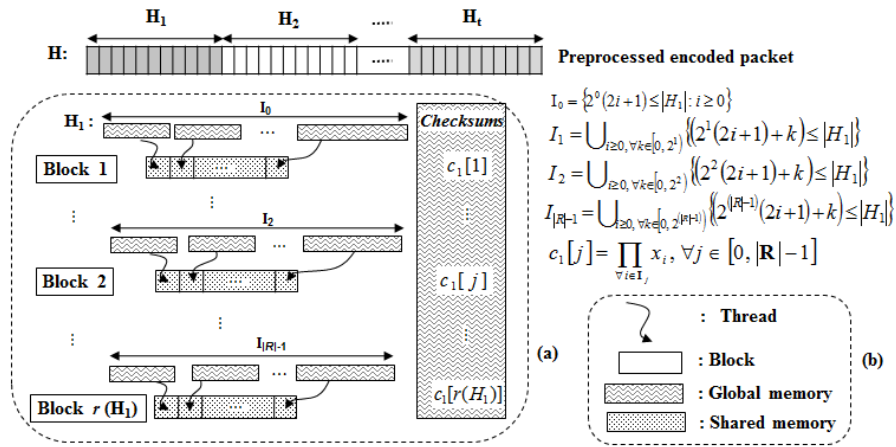}
\caption{Kernel \textit{checksum} to calculate checksum on packet segment $\textbf{H}_{1}$.} \label{fig:numbytes1}
\end{center}
\end{figure*}


\vspace{10pt}
\section{Experimental Results and Analysis}
\label{sec:experimental}

To execute the CPU code, we use a machine running on Windows 7 (32 bit) with a 4-core 3.40GHz Intel processor that utilizes 8GB main memory. We evaluate the performance of the proposed GPU implementation on a 1.62GHz NVIDIA GeForce GTX 560 GPU with seven streaming multiprocessors (SM) and 1GB of main memory, where the GPU employed 336 processing elements and utilized 49,152 bytes of shared memory per SM. Furthermore, the maximum threads per block are 1,024 and the warp size is 32 threads. In this section, the execution times of the GPU-based and CPU-based Hamming decoders are compared. 

\vspace{10pt}
\subsection{Execution Time}
\label{sec:executiontime}

Figures 7 and 8 show the consolidated execution time according to packet size, \textbf{M}, and error tolerance, \textit{t}, for sequential and parallel Hamming decoding on the CPU and GPU, respectively. The packet length ranges from 400 bytes to 2000 bytes (x-axis) and the error tolerance is tested from 2 bits up to 6 bits (y-axis); the decoding time in milliseconds (\textit{ms}) is shown on the z-axis.

\begin{figure}[h]
\begin{center}
\epsfxsize=7.3cm \leavevmode\epsfbox{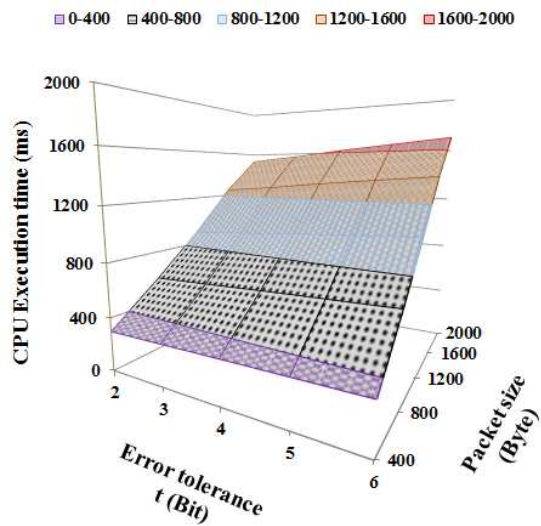}
\caption{Execution time of the CPU-based decoder.} \label{fig:numbytes1}
\end{center}
\end{figure}

 Execution time generally increases with packet size for the following two reasons. First, longer messages have a greater number of checksum bits attached to the message. Second, the code used to calculate redundant information is also lengthened, increasing the number of XOR operations required. In addition, the computational time of CPU implementation is proportionally influenced by the Hamming code length of the packet. In contrast, the time for the GPU to decode packages remains relatively constant regardless of packet size.

There is a gradual increase in execution time as error tolerance increases for the CPU, while execution time of the GPU decreases as error tolerance, \textit{t}, increases. The increase in \textit{t} implies that a large number of bits in the packet are corrupted, because the transmission medium is erroneous. In these situations, the decoder splits away the packet into a higher number of segments, as shown in Figure 1. The CPU-based approach finishes these segments in a sequential manner, leading to an increase in decoding time. In contrast, the GPU deals with the segments in parallel and thus mitigates the increase in \textit{t} with error tolerance. In addition, greater partitioning of a packet results in a smaller 
\\
\begin{figure}[h]
\begin{center}
\epsfxsize=7.4cm \leavevmode\epsfbox{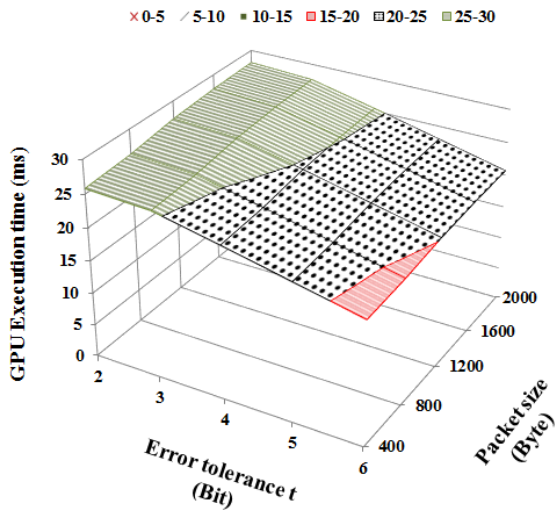}
\caption{Execution time of the GPU-based decoder.} \label{fig:numbytes1}
\end{center}
\end{figure}
\\
segment size, which enables the GPU to achieve a faster execution time than the CPU.

Overall, the GPU outperforms the CPU in terms of time required to decode various packet sizes and error tolerance, yielding a tremendous improvement in execution time. We attribute this to the massively parallel design of SP, DE, and ME of the GPU. Detailed speedup information is summarized in Table 1; the maximum speedup gained by the GPU over the CPU is 99$\times$.

\begin{table}[h]
\caption{Speedup of GPU over CPU.}
\label{tab:speedup}
\begin{center}
\begin{tabular}{cccccc}\hline
Packet size & t=2 & t=3 & t=4 & t=5 & t=6 \\ \hline 
M=400 & 13$\times$ &  14$\times$ & 16$\times$ & 18$\times$ & 21$\times$  \\  
M=800 & 26$\times$ &  27$\times$ & 30$\times$ & 35$\times$ & 40$\times$  \\ 
M=1200 & 38$\times$ & 40$\times$ & 45$\times$ & 51$\times$ & 60$\times$  \\ 
M=1600 & 52$\times$ & 54$\times$ & 61$\times$ & 70$\times$ & 81$\times$  \\ \hline
\end{tabular}
\end{center}
\end{table}

\vspace{0pt}
\section{Conclusions}
\label{sec:experimental}

In this paper, we proposed a computationally efficient GPU implementation of a Hamming code decoder for faster error recovery in data communication networks. We compared the performance of the proposed GPU approach with an equivalent sequential approach on a traditional CPU. The GPU-based implementation strongly outperformed the CPU-based sequential approach in terms of execution time, yielding a 99$\times$ speedup. These results indicate that the proposed GPU approach is suitable for application in time-sensitive and high-speed wired and wireless communication systems.

\bibliographystyle{jcn}


\epsfysize=3.2cm
\begin{biography}{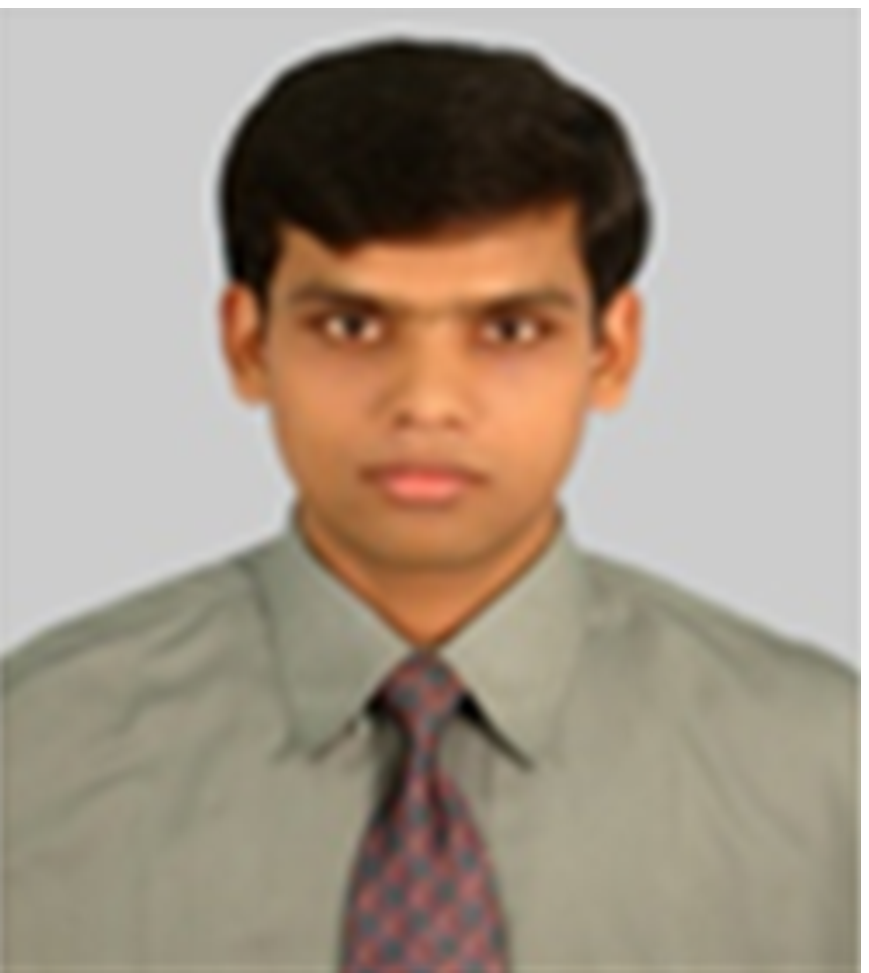}{Md Shohidul Islam}
received a B.S. degree in computer science and engineering from Rajshahi University of Engineering and Technology, Rajshahi, Bangladesh, in 2007, and an M.S. in computer engineering from the University of Ulsan, Ulsan, South Korea, in 2014. His current research interests include high-speed error coding, GPU computing, software defined radio (SDR), parallel computing, and the performance evaluation of many-core processors for application-specific SoC design.
\end{biography}

\epsfysize=3.2cm
\begin{biography}{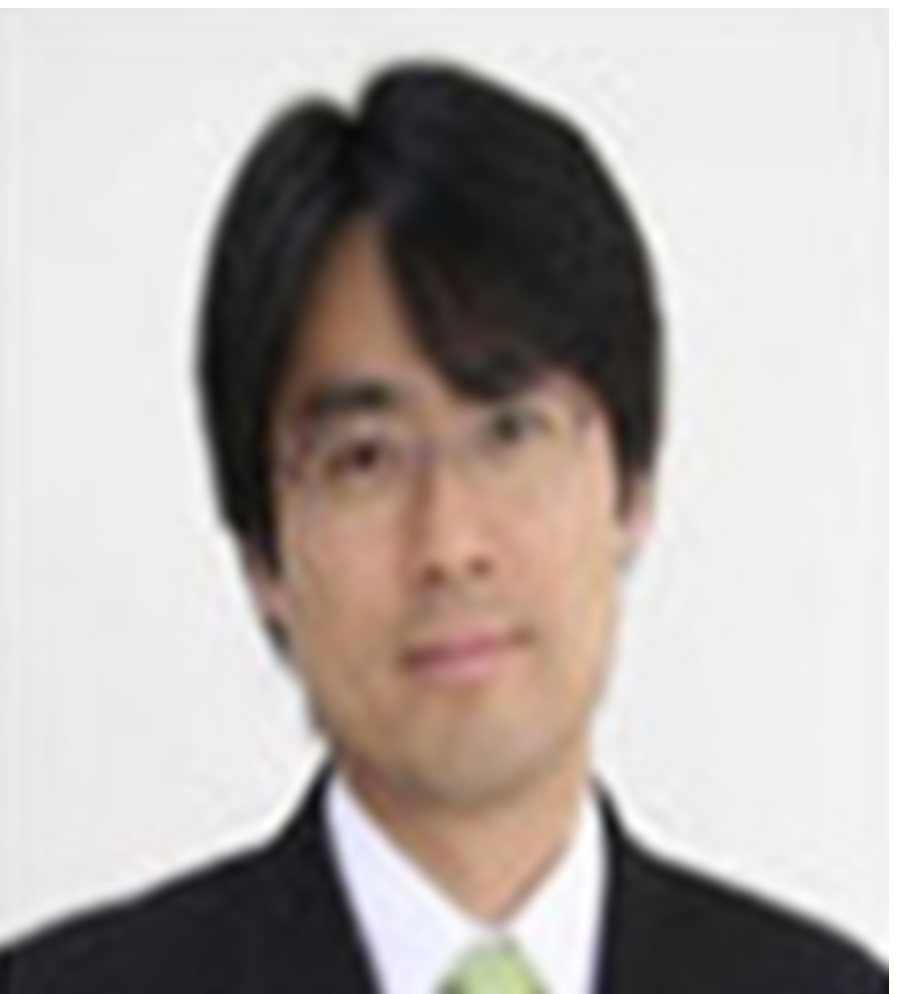}{Cheol-Hong Kim} received a BS, a MS, and a PhD in computer engineering from Seoul National University, Seoul, Korea, in 1998 2000, and 2006, respectively. He is an Associate Professor of Electronics and Computer Engineering at Chonnam National University, Gwangju, Korea. His research interests include multicore architecture and embedded systems.
\end{biography}

\newpage

\epsfysize=3.2cm
\begin{biography}{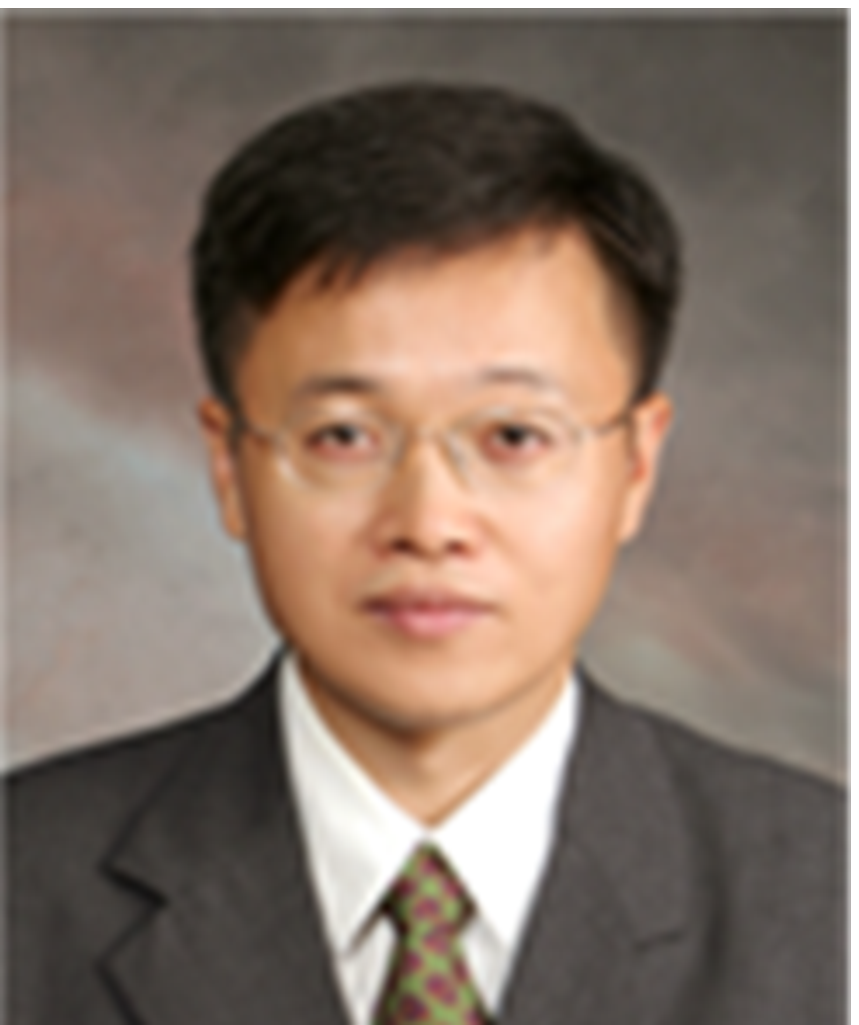}{Jong-Myon Kim} received a B.S. degree in electrical engineering from Myongji University, Yongin, South Korea, in 1995, an M.S. in electrical and computer engineering from the University of Florida, Gainesville, in 2000, and a Ph.D. in electrical and computer engineering from the Georgia Institute of Technology, Atlanta, in 2005. He is an Associate Professor of Electrical Engineering at the University of Ulsan, Ulsan, South Korea. His research interests include GPU computing, multimedia specific processor architecture, parallel processing, and embedded systems. He is a member of IEEE and IEICE.

\end{biography}

\end{document}